# Quantum Interferometry and Correlated Two-Electron Wave-Packet Observation in Helium

Christian Ott, Andreas Kaldun, Philipp Raith, Kristina Meyer, Martin Laux, Yizhu Zhang, Steffen Hagstotz, Thomas Ding, Robert Heck, and Thomas Pfeifer

Max-Planck Institut für Kernphysik, Saupfercheckweg 1, 69117 Heidelberg, Germany
Center for Quantum Dynamics, Ruprecht-Karls-Universität Heidelberg, 69120 Heidelberg, Germany

**The concerted motion of two or more bound electrons governs atomic[1] and molecular[2,3] non-equilibrium processes and chemical reactions. It is thus a long-standing scientific dream to measure the dynamics of two bound correlated electrons in the quantum regime. Quantum wave packets were previously observed for single-active electrons on their natural attosecond timescales[4-6]. However, at least two active electrons and a nucleus are required to address the quantum three-body problem[7]. This situation is realized in the helium atom[7-16], but direct time-resolved observation of two-electron wave-packet motion remained an unaccomplished challenge. Here, we measure a 1.2-femtosecond quantum beating among low-lying doubly-excited states in helium to evidence a correlated two-electron wave packet. Our experimental method combines attosecond transient-absorption spectroscopy[4,17,18] at unprecedented high spectral resolution (20 meV near 60 eV) with an intensity-tuneable visible laser field to couple[9-11] the quantum states from the perturbative to the strong-coupling regime. This multi-dimensional transient-coupling scheme reveals an inversion of the characteristic Fano[8,19] line shapes for a range of doubly-excited states[12]. Employing Fano-type autoionization as a natural quantum interferometer, a dynamical phase shift by laser coupling to the $N=2$ continuum is postulated and experimentally quantified. This phase maps a transition from effectively single-active-electron to two-electron dynamics as the electron–electron interaction increases in lower-lying quantum states. In the future, such experiments will provide benchmark data for testing dynamical few-body quantum theory. They will boost our understanding of chemically and biologically important metastable electronic transition states and their dynamics on attosecond time scales.**

Electrons are bound to atoms and molecules by the Coulomb force of the nuclei. Moving between atoms, they form the basis of the molecular bond. The same Coulombic force, however, acts repulsively between the electrons. This electron–electron interaction represents a major challenge in the understanding and modelling of atomic and molecular states, their structure and in particular their dynamics[2,3,20]. The helium atom is an ideal system to study two-electron dynamics[7-16] for which exact analytical solutions do not exist. This is in contrast to single-active electron (e.g. hydrogen-like) systems, which are often used as approximate pictures to describe two- or multi-electron systems, however always raising fundamental questions such as: In which regime is the single-active electron picture valid, and where does it break down? To find answers to these questions, we focus on the $^1P$ sp$_{2,n+}$ series[12] of doubly-excited states in helium. They are excited by a single-photon transition from the $^1S$ $1s^2$ ground state by the promotion of both electrons to at least principal quantum number $n$=2. These states are metastable and autoionize due to electron–electron interaction. Non-Lorentzian asymmetric line shapes have been spectroscopically observed in the 1930s[19] and attributed, 50 years ago by Ugo Fano[8], to quantum interference (coupling) of bound states with a continuum (Figure 1c,d). The coupling is described by the configuration-interaction $V_{CI}$ with the single-ionization continuum $|1s,\varepsilon p\rangle$, where one electron is in the 1s ground state and the other one is in the continuum with kinetic energy $\varepsilon$. The lifetimes of the bound states are governed by the magnitude of $V_{CI}$, ranging between 17 fs for the 2s2p (sp$_{2,2}$) [21] and several hundreds of femtoseconds for the highest-lying sp$_{2,n+}$ states[12]. As their lifetimes are that short and energy-level spacings are on the order of several eV, their couplings and dynamics must be measured using ultrashort pulses. Previous time-resolved experiments on these He states observed light-induced modification of absorption profiles[17,18] and used attosecond streak-field spectroscopy[21] to measure the 2s2p autoionization life time. A 1.2-fs two-electron wave packet formed by superposition of the lowest-lying autoionising states was recently predicted theoretically[15]. However, experimental observation of such two-electron motion deep in the quantum regime (low-$n$ states) failed so far due to the technical challenge of achieving both high temporal and high spectral resolution to recover the corresponding fast and spectrally-narrow (Fano) features.

Our experimental method (Figure 1a,b) augments existing attosecond transient-absorption approaches by adding high-spectral-resolution capability with a soft-x-ray (SXR) flat-field grating spectrometer. It allows the parallel measurement of *spectrally narrow* absorption lines imprinted on an attosecond-pulsed *broad-band* SXR spectrum in the presence of a visible (VIS) laser field. The VIS laser couples the two-electron excited states (Figure 1c). The time-delay between VIS and SXR fields and the intensity of the VIS field are varied as experimental parameters to create a novel multi-dimensional transient-coupling scheme.

Absorption spectra for the variation of VIS-SXR time-delay at constant VIS intensity are shown in Figure 2. Besides a shifting, splitting and broadening of the main absorption lines during VIS and attosecond SXR pulse temporal overlap (previously reported in inner-valence excitation of argon[18]) additional fine-scale spectral structures as well as single-femtosecond temporal oscillations throughout the spectrum are resolved. The two lowest-lying states 2s2p and sp$_{2,3+}$ are most strongly modulated as a function of time delay, revealing the presence of ultrafast coherent two-electron dynamics — a two-electron wave packet — with a period of ~1.2 fs.

To quantitatively analyze the wave-packet dynamics, we performed a Fourier transformation along the time-delay axis (Figure 3), creating a two-dimensional spectrogram. A similar modulation frequency $\Delta\omega$ on both the 2s2p ($\hbar\Delta\omega_1$=3.51±0.18 eV) and sp$_{2,3+}$ ($\hbar\Delta\omega_2$=3.36±0.18 eV) lines is observed, in agreement with the 3.51 eV energy difference $\Delta E=\hbar\Delta\omega$ between the 2s2p and sp$_{2,3+}$ states[12], evidencing the correlated bound two-electron wave packet[15] $|\Psi(t)\rangle \propto (|2s2p\rangle + a\exp(-it\Delta E/\hbar)|sp_{2,3+}\rangle)$ with the complex mixing parameter $a$. From our theoretical modelling (see SI Section 3), we understand the observed 1.2 fs wave-packet beating as a consequence of the VIS-induced two-photon dipole coupling of the 2s2p and the sp$_{2,3+}$ states proceeding via the energetically intermediate and spectroscopically dark 2p$^2$ state (at

62.06 eV, illustration of energy levels see Figure 1). The 2s2p↔2p$^2$ transition alone was previously used to control the soft-x-ray transmission of helium[11,17,21].

After understanding and validating our experimental transient-coupling method by observing the wave packet, we proceed to expanding its applicability for general two-electron quantum interferometry. The electric field strength of the VIS pulse is an important parameter: it controls the coupling strength between states. Continuous variation of the VIS pulse intensity thus opens a 3$^{rd}$ dimension in our spectroscopy method, in addition to time delay and photon energy. Further important questions can now be addressed: How do two-electron transition states respond to field strengths ranging from perturbative to the strong-field limit? What is the dynamics and fate of doubly-excited states at and before the onset of ionization? What is the validity range of commonly-used[17,21,22] single-active electron pictures for strong-field ionization of two-electron systems?

Figure 4 shows the results of VIS laser intensity tuning in our experiment at a fixed time delay of ~5 fs (OD spectra were averaged over one VIS optical cycle). It continuously maps the transition from the unperturbed to the strong-coupling regime of discrete doubly-excited states that is evident near 60 eV. At the highest intensities, a total of three lines is visible, confirming the contribution of VIS-dressed states originating from 2s2p, 2p$^2$, and sp$_{2,3+}$ by the absorption of effectively zero, one, and two VIS photons, respectively. Furthermore, all states are observed to resist the electric fields of the laser far beyond the field strengths resulting in classical detachment of the outermost electron by overcoming the attractive nuclear Coulomb force (over-the-barrier ionization[22]). This resilience to laser ionization appears particularly counterintuitive given the autoionizing nature of these states, for which ionization proceeds even in the absence of a laser field.

Due to our high spectral resolution, Figure 4 also reveals characteristic changes in the shapes of the Fano line profiles. For all states sp$_{2,n+}$ with $n \geq 3$ and starting at lowest intensities, the Fano profile starts to inverse at intensities larger than ~3×10$^{12}$ W/cm$^2$. According to Fano's theory, the autoionization cross section is given by[8]

$$\sigma(E = \hbar\omega) = \frac{(q+\Delta)^2}{1+\Delta^2}; \quad \Delta = \frac{E - E_0}{\Gamma},$$  **Eq. 1**

with $E_0$ and $\Gamma$ the resonance energy and width, respectively. The line-shape inversion could be described by the negation of the Fano $q$ parameter, described as

$$q = \frac{\langle \text{sp}_{2,n+} | T | 1s^2 \rangle}{\pi V_{CI}^* \langle 1s, \varepsilon p | T | 1s^2 \rangle}.$$  **Eq. 2**

We can thus interpret the line-shape inversion as the negation of at least one of the following quantities: the matrix elements $T_n = \langle \text{sp}_{2,n+} | T | 1s^2 \rangle$ and $T_\varepsilon = \langle 1s, \varepsilon p | T | 1s^2 \rangle$, describing the transition $T_n$ from the helium ground state to the bound states and the continuum, respectively, and the configuration interaction $V_{CI}$. The continuum transition $T_\varepsilon$ occurs on the time scale of the attosecond pulse and is thus unaffected by the VIS pulse, arriving ~5 fs later. The VIS pulse is much shorter (~7 fs) than the lifetimes of the highly-excited states (>100 fs). Thus, it cannot change $V_{CI}$ for the entire course of autoionization.

Here we hypothesize that the laser can, however, effectively modify the transition $T_n$ into these states by imprinting an additional phase. What could be the mechanism for this phase shift? In single-electron pictures, loosely bound electrons would behave as free electrons in their response to laser fields. As such, they experience a dynamical phase shift by their additional kinetic quiver motion (ponderomotive energy) in the oscillating electric field of the laser:

$$\Delta\varphi = \int_{-\infty}^{\infty} \tfrac{1}{2} m_e v^2(t') dt' \qquad \text{Eq. 3}$$

With $v(t) = -\dfrac{e}{m_e}\int_{-\infty}^{t} E(t')dt'$, the velocity of the electrons. For states $|sp_{2,n+}\rangle$ with $n>4$, we assume this additional phase to be mostly accumulated by one, the outermost, electron. However, it is imprinted on the entire bound correlated two-electron state $|sp_{2,n+}\rangle \sim |2s,n\text{p}\rangle + |2\text{p},n\text{s}\rangle$, for which the outer electron, interestingly, is in a superposition of $n$s and $n$p, entangled with the innermost 2p and 2s states, respectively. For our pulse duration of 7 fs and from Eq.(3), we expect a phase shift of $\pi$ near an intensity of $\sim 5\times 10^{12}$ W/cm$^2$. This is in good agreement with the experiment: for the high-lying states with $n>4$ Fano line inversion is observed at intensities larger than $3\times 10^{12}$ W/cm$^2$ (Figure 4b).

The VIS pulse is thus able to quantitatively control one arm (the autoionizing transition state) of the natural two-electron quantum interferometer by coherently coupling one of the two electrons to the N=2 continuum (see Figure 1c). Interestingly, this separation of the correlated/entangled $sp_{2,n+}$ states into an active and a spectator electron during the interaction with the laser, and back to the $sp_{2,n+}$ after the laser interaction, does not destroy the coherence. This is evidenced by the Fano line shapes that remain fully intact, visible, and narrow even at high intensities and just shift their phase.

The Fano profile is not fully inverted Figure 4b) for the lower-lying states $|sp_{2,n+}\rangle$ with $n\leq 4$, as they are less suitably described by the quasi-free single-electron picture. This $n$-dependence marks the transition from the effectively single-electron, quasi-classical dynamics regime ($n>\sim 4$) to the two-electron quantum dynamics regime ($n\leq 4$). The state-dependent magnitude of these phase changes will allow further insight into the coupling among two electrons and how they, collectively or cooperatively, acquire dynamical phases under external perturbation.

Having state-resolved access to and control over the full quantum information — amplitude and phase — for two-electron-excited states, more fundamental questions can now be addressed: What is the shape of such two-electron correlated wavefunctions in external fields? To what extent can it be controlled? Do several electrons ionize independently or cooperatively? The answers carry important consequences for current scientific dreams, e.g. to create synthetic atomic quantum systems and exotic molecules beyond the reaches of traditional thermodynamic chemistry by ultrafast temporally tailored light fields[23], and for applications such as solar-energy conversion, coherent excitation transfer in biosystems (photosynthesis, FRET), (quantum-)information processing and (nano-)material machining by laser-induced plasma.

**Methods Summary**

A few-cycle amplified laser system (Femtolasers Compact Pro) with hollow-fibre and chirped-mirror compression stages delivers 5-7 fs pulses at a 4 kHz repetition rate. These pulses are coupled into our home-built attosecond interferometry vacuum beamline (Figure 1 and SI Section 1) consisting of a SXR source chamber, refocusing and time-delay chamber and a target chamber followed by a high-resolution ($\Delta\lambda$ = 0.01 nm, corresponding to $\Delta E$ = 20 meV near ~60 eV) SXR spectrometer. Isolated and few-pulse trains of <500 as duration SXR pulses are generated by high-order harmonic generation (HHG) in a gas cell filled with neon (~100 mbar). The generated SXR pulses then co-propagate with their intense driver pulse and impinge at 15-degree grazing-incidence angle on a two-component split mirror. The SXR pulses reflect off a smaller inner mirror (centred on the optical axis) while the more divergent VIS pulses are reflected by a larger outer mirror surrounding the inner mirror. The inner mirror can be translated parallel to its surface normal to create high-precision (10 as) time delays between the SXR attosecond and VIS femtosecond pulses. Inserting an aluminium filter and a polymer membrane with a centre hole, we create an annular femtosecond VIS laser beam that is fully separable in time (delay) and space from the inner attosecond SXR beam. These pulses are then focused into a second gas cell of 3 mm length — providing the helium used in this study at a pressure of 100 mbar — using a toroidal mirror (focal length 0.35 m) under 15-degree grazing angle. The VIS pulse intensity is tuned by a motor-controlled iris aperture before focusing into the helium target. The focus in the gas cell also represents the entrance slit of our flat-field (variable-line-space grating) spectrometer, recording the absorption spectrum in the range between 20 and 120 eV.

**Acknowledgements** We thank E. Lindroth for calculating the dipole moment $\langle 2p^2|\mathbf{r}|sp_{2,3+}\rangle$, and also A. Voitkiv, J. Madroñero, Z.-H. Loh, and R. Moshammer for helpful discussions. We acknowledge financial support by the Max-Planck Research Group Program of the Max-Planck Gesellschaft (MPG).

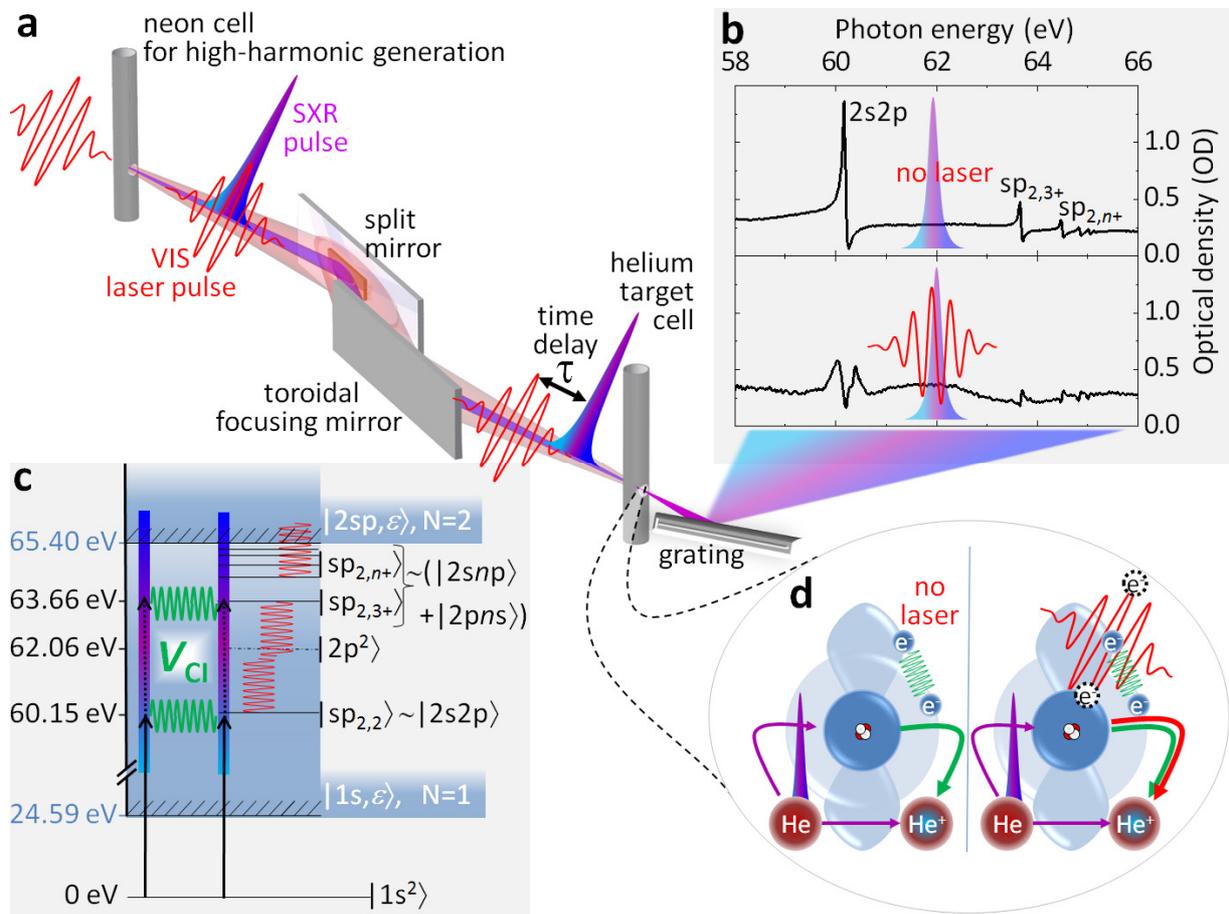

**Figure 1 | Illustration of the experimental setup, data, and microscopic mechanisms in helium. a,** Few-cycle (7 fs) VIS laser pulses (730 nm) are focused into a neon gas cell for partial attosecond-pulse conversion, providing a continuous coherent probe-light spectrum throughout the SXR range. The time delay between co-propagating VIS and SXR pulses is controlled by a split(inlay)-mirror stage. Both pulses transit the helium gas target and enter the high-resolution spectrometer. **b,** Absorption spectra without (upper) and in the presence of the VIS laser pulse (lower), throughout the region of the $|sp_{2,n}\rangle$ doubly-excited states. **c,** Helium level diagram: The $|sp_{2,n}\rangle$ states couple to the $|1s,\varepsilon\rangle$ continuum by configuration interaction $V_{CI}$ (green springs), resulting in characteristic Fano spectral line shapes[8] caused by quantum interference. State-resolved phase information can be measured by coupling this natural quantum interferometer to the controlling VIS laser field (red springs). **d,** microscopic mechanisms: The SXR pulses can either directly ionize He to He$^+$, or excite both electrons into a metastable transition state, which then decays by configuration interaction $V_{CI}$ into He$^+$, quantum interfering with the direct ionization process (left, natural process). If a laser field is present (right), it shifts the phase of one arm of this natural interferometer—the two-electron transition state—modifying the Fano line shapes detected in the transmitted absorption spectrum thus serving as a measure of this quantum phase.

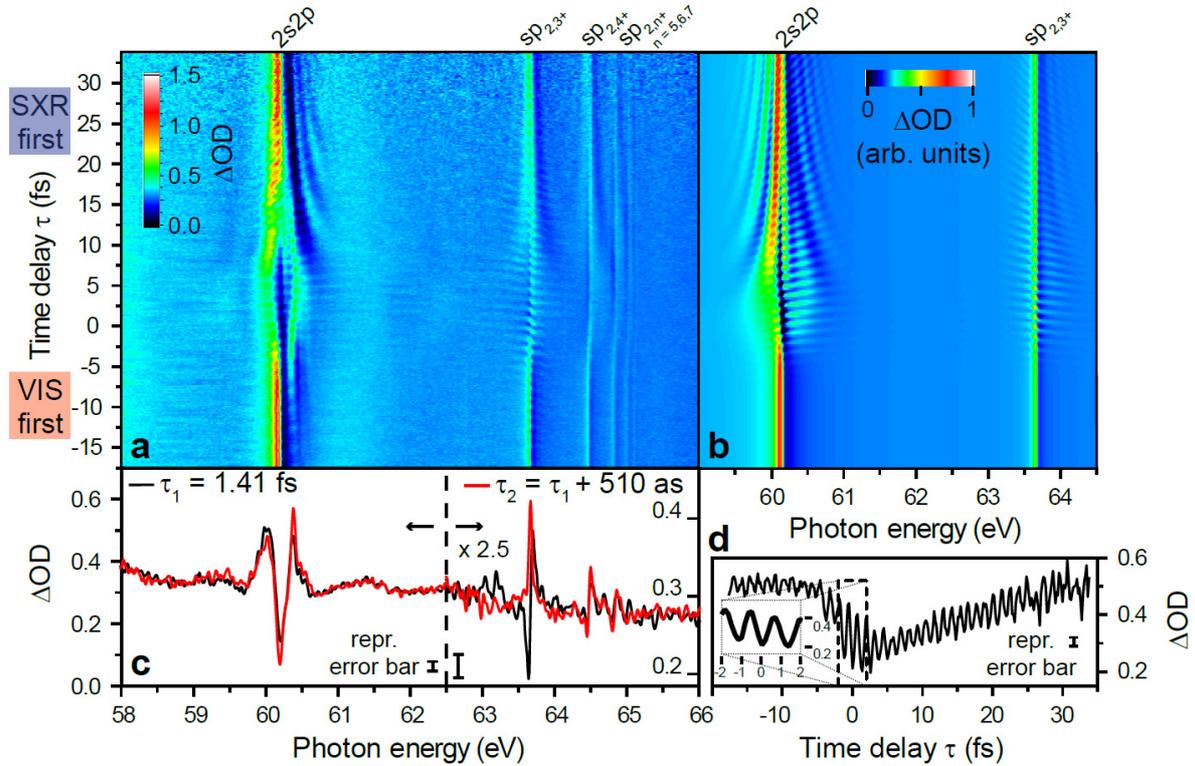

**Figure 2 | Observation of two-electron wave packet dynamics in helium. a, b,** Absorption of SXR light in helium versus time delay between the VIS ($3.3 \times 10^{12}$ W/cm$^2$ intensity) coupling field and the SXR attosecond probe pulses. Experiment (**a**), and simulations (**b**) based on a four-level+continuum quantum model (see SI Section 3). At early times, the well-known Fano absorption line profile[8] is measured for several autoionizing states up to sp$_{2,7+}$. In the region near temporal overlap and at late times, the absorption spectrum is strongly modified. At slightly positive time delays (SXR pulse immediately preceding the VIS pulse) a clear signature of Autler-Townes splitting of the 2s2p resonance with the energetically repelling 2p$^2$ dressed state is measured at ~60 eV, confirming the strong-coupling regime of autoionizing states and multiple Rabi cycling among these two states. A 1.2-fs oscillation evidences the creation of a two-electron-excited wave packet, arising from the VIS laser coupling of 2s2p, 2p$^2$, and sp$_{2,3+}$ states. **c,** Lineout vs. photon energy at different time delays. As a consequence of the discrete-state coupling, the correlated two-electron state Fano resonance sp$_{2,3+}$ at 63.66 eV depends sensitively (0.5 fs) on the time-delay. **d,** Lineout of the OD signal at the sp$_{2,3+}$ resonance of 63.66 eV vs. time delay. The 1.2 fs time-delay modulation is clearly visible; the inset shows a magnified view near zero time delay.

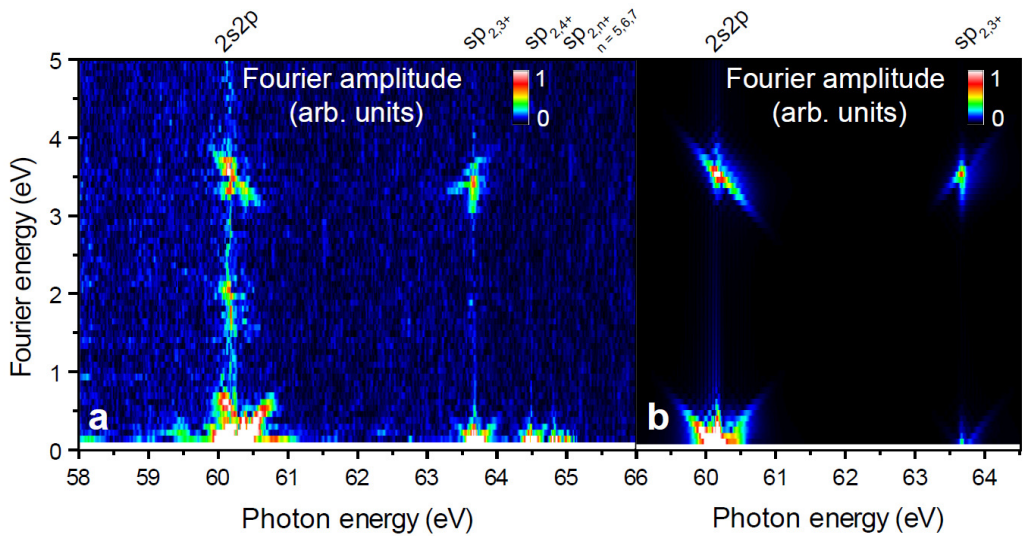

**Figure 3 | Two-dimensional spectrogram (2dS), revealing coupling of doubly-excited states:
a, b,** The 2dS is obtained from the data shown in Fig. 2 by Fourier transformation along the time-delay axis for the experiment (**a**), and the simulation results (**b**). The experimental observation of the fast frequency 3.51 eV/$h$=0.85 PHz) component in both 2s2p and $sp_{2,3+}$, but not in other configuration states ($sp_{2,4+}$ and higher), quantitatively proves the coherent coupling of these low-lying autoionizing quantum states by the VIS field. The diagonal lines originating at each 2dS peak (both at 3.5 and 0 eV) are due to off-resonant interference of the natural and laser-dressed two-electron dipole responses and further confirm the coherent coupling and wave-packet nature of these states. The component near 2 eV in the experiment is explained by a small leakage of the fundamental laser field into the time-delayed SXR beam. The SXR-VIS 2dS measurement can also be regarded as a stepping stone towards 2d x-ray spectroscopy as theoretically suggested[24], to be implemented at x-ray Free-Electron Laser (FEL) facilities in the near future.

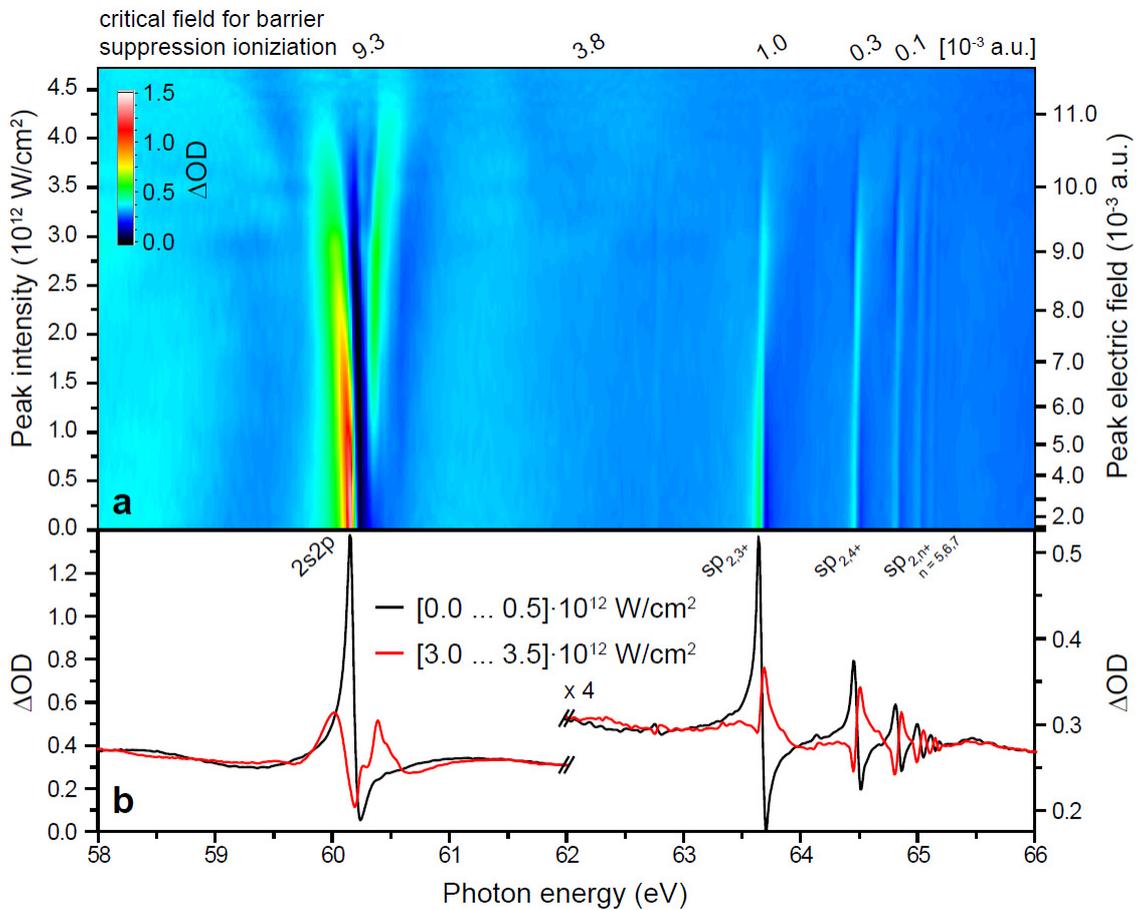

**Figure 4 | Intensity-dependent strong coupling and observation of dynamical phase shifts for two-electron states. a,** SXR absorption spectra at a time-delay of (5.2+/–1.2) fs for increasing VIS coupling intensities. Near 60 eV we continuously follow the transition from the unperturbed to the two-electron strong-coupling regime of the 2s2p with $2p^2$ and $sp_{2,3+}$: at the highest intensities, a total of three mutually repelling absorption lines is present. Most interestingly, for the higher-lying excited states, we can measure the modification and inversion of the Fano quantum-interference profiles, revealing a phase change of the corresponding two-electron wavefunctions. **b,** Lineouts at different intensities. For the highest-lying states, laser-induced inversion of Fano line shapes occurs by a quasi-classical single-active electron process (see text) while for lower states quantum two-electron dynamics becomes more dominant resulting in different line-shape modifications. All states are fully visible and stable well above critical fields corresponding to classical (Coulomb-barrier-suppressed) ionization. However, the reason why basically *all* states seem to collectively and abruptly vanish at an identical critical intensity of $\sim 4 \times 10^{12}$ W/cm² is not yet understood within existing theory. Error bars in (b) are given by approximately twice the line thickness.